\documentclass[aps,prl,amsmath,twocolumn,superscriptaddress,showpacs]{revtex4-1}

\usepackage{graphicx}
\usepackage{color}
\usepackage[normalem]{ulem}

\begin{document}

\title{Using controlled disorder to probe the interplay between \\charge order and superconductivity in NbSe$_2$}

\author{Kyuil~Cho}
\affiliation{Ames Laboratory, Ames, IA 50011, USA}
\affiliation{Department of Physics $\&$ Astronomy, Iowa State University, Ames, IA 50011, USA}

\author{M.~Ko\'nczykowski}
\affiliation{Laboratoire des Solides Irradi\'es, {\'E}cole Polytechnique, CNRS, CEA, Universit\'e Paris-Saclay, F-91128 Palaiseau, France}

\author{S.~Teknowijoyo}
\affiliation{Ames Laboratory, Ames, IA 50011, USA}
\affiliation{Department of Physics $\&$ Astronomy, Iowa State University, Ames, IA 50011, USA}

\author{M.~A.~Tanatar}
\affiliation{Ames Laboratory, Ames, IA 50011, USA}
\affiliation{Department of Physics $\&$ Astronomy, Iowa State University, Ames, IA 50011, USA}

\author{J.~Guss}
\affiliation{Ames Laboratory, Ames, IA 50011, USA}

\author{P.~B.~Gartin}
\affiliation{Ames Laboratory, Ames, IA 50011, USA}
\affiliation{Department of Physics $\&$ Astronomy, Iowa State University, Ames, IA 50011, USA}

\author{J.~Wilde}
\affiliation{Ames Laboratory, Ames, IA 50011, USA}
\affiliation{Department of Physics $\&$ Astronomy, Iowa State University, Ames, IA 50011, USA}

\author{A.~Kreyssig}
\affiliation{Ames Laboratory, Ames, IA 50011, USA}
\affiliation{Department of Physics $\&$ Astronomy, Iowa State University, Ames, IA 50011, USA}

\author{R.~McQueeney}
\affiliation{Ames Laboratory, Ames, IA 50011, USA}
\affiliation{Department of Physics $\&$ Astronomy, Iowa State University, Ames, IA 50011, USA}

\author{A.~Goldman}
\affiliation{Ames Laboratory, Ames, IA 50011, USA}
\affiliation{Department of Physics $\&$ Astronomy, Iowa State University, Ames, IA 50011, USA}

\author{V.~ Mishra}
\affiliation{Computer Science and Mathematics Division, Oak Ridge National Laboratory, Oak Ridge, TN 37831, USA}

\author{P.~J.~Hirschfeld}
\affiliation{Department of Physics, University of Florida, Gainesville, FL 32611, USA}

\author{R.~Prozorov}
\email[Corresponding author: ]{prozorov@ameslab.gov}
\affiliation{Ames Laboratory, Ames, IA 50011, USA}
\affiliation{Department of Physics $\&$ Astronomy, Iowa State University, Ames, IA 50011, USA}

\date{Submitted: 11 October 2017; Published: Nat Comm \textbf{9}, 2786 (18 July 2018)}

\begin{abstract}
The interplay between superconductivity and charge density waves (CDW) in $H$-NbSe2 is  not fully understood despite decades of study.  Artificially introduced disorder can tip the delicate balance between two competing forms of long-range order, and reveal the underlying interactions that give rise to them.  Here we introduce disorders by electron irradiation and measure in-plane resistivity, Hall resistivity, X-ray scattering, and London penetration depth. With increasing disorder, $T_{\textrm{c}}$ varies nonmonotonically, whereas $T_{\textrm{CDW}}$ monotonically decreases and becomes unresolvable above a critical irradiation dose where $T_{\textrm{c}}$  drops sharply. Our results imply that CDW order initially competes with superconductivity, but eventually assists it. We argue that at the transition where the long-range CDW order disappears, the cooperation with superconductivity is dramatically suppressed. X-ray scattering and Hall resistivity
measurements reveal that the short-range CDW survives above the transition.   Superconductivity persists to much higher dose levels, consistent with fully gapped superconductivity and moderate interband pairing.
\end{abstract}
\maketitle

The interplay between superconductivity (SC) and density wave orders has been a central issue in high temperature superconductors such as cuprates and iron-based superconductors \cite{Dai2015RMP_revew_FeSCs_Neutron}.
 {The recent discovery of} a charge density wave (CDW) phase in the middle of the pseudogap region of  {cuprates \cite{Ghiringhelli2012Science_YBCO_CDW, Chang2012NatPhys_YBCO_CDW,Achkar2012,Comin390,Neto2014,Tabis2014} has} revitalized
interest in the interplay between CDW and  superconducting states in other {unconventional} superconductors, such as the layered transition-metal dichalcogenides, {in particular well-studied 2$H$-NbSe$_2$  }\cite{GabovichVoitenko2001SST_SC_CDW_SDW_review, WilsonSalvo1975AdvPhys_dichalcogenides, Moncton1975PRL_NbSe2_Neutron, HARPER1975PhysLettsA_NbSe2_HeatCapacity_CDW}.  This system has fascinated investigators for decades
due to microscopic coexistence of CDW ($T_{\textrm{CDW}} = 33$ K) and SC ($T_{\textrm{c}}=7.2$ K) states \cite{GuillamonSuderow2008PRB_NbSe2_STM_modulation, SooryakumarKlein1980PRL_NbSe2_Raman_CDW}. The coupling between the two long-range orders is apparently responsible for the observability of the elusive Higgs bosonic amplitude  mode of the superconductor \cite{VarmaLittlewood81,VarmaLittlewood82}, discovered by Raman scattering on 2$H$-NbSe$_2$ \cite{Measson2014, Grasset2018PRB_NbSe2_Raman}.

Despite intense effort, however, a key question regarding the nature of the coupling of the two orders in this system is still under debate. In recent years, the conventional weak-coupling picture where CDW and SC compete for parts of the Fermi surface has been challenged by the realization that the electron-phonon coupling is very strong due to the
two-dimensional confinement of the Nb layer \cite{Moncton1975PRL_NbSe2_Neutron, Inglesfield1980JPC_dichalcogenide, VarmaSimons1983PRL_CDW_theory,  Weber2011PRL_NbSe2_CDW_Phonon, Flicker2015NatComm_NbSe2_CDW}.  In such a situation, the usual mean field picture of a charge density wave order  with rigid amplitude and phase disappearing at $T_{\textrm{CDW}}$ may no longer be valid, since the short-range CDW order together with a gap in the electronic spectrum have been observed outside the long-range ordered phase \cite{Chatterjee2015}.

The situation in 2$H$-NbSe$_2$ is complicated by the complex electronic bandstructure of this material and apparent multiband superconductivity with two effective gaps \cite{FletcherProzorov2007PRL_NbSe2_TDR}.  Different superconducting gaps on different Fermi surface sheets were inferred from angle-resolved photoemission spectroscopy (ARPES) by Yokoya {\it et al.} \cite{Yokoya2001Science} and thermal conductivity measurements by Boaknin {\it et al.} \cite{Boaknin2003PRL_NbSe2_Multiband}. 
Kiss {\it et al.} proposed that the CDW
actually boosts the superconductivity, based on the correlation with the largest electron-phonon coupling and lowest Fermi velocities at the same $\bf l$-points \cite{Kiss2007NatPhys_CDW_SC_ARPES}. 
Borisenko {\it et al.} observed Fermi arcs, suggesting that the CDW inhibits the formation of superconducting order by gapping the nested portion of Fermi surface
\cite{Borisenko2009PRL_NbSe2_ARPES}.

The pressure dependence of $T_{\textrm{CDW}}$ and $T_{\textrm{c}}$ is another way to study the interplay between the CDW and superconductivity. Leroux {\it et al.} suggest that the pressure and temperature dependence of the phonon dispersion, observed by  inelastic X-ray scattering, support insensitivity of $T_{\textrm{c}}$ to the CDW transition \cite{LerouxRodiere2015PRB_NbSe2_Pressure}. However, Feng {\it et al.} reported a broad regime of order parameter fluctuations in X-ray diffraction (at $T$ = 3.5 K), and attributed it to the presence of a CDW quantum critical point (P$_{CDW}$ = 4.6 GPa) buried beneath the superconducting dome \cite{Feng2012PNAS_CDW_QCP}. They also claimed that this is partially consistent with the increasing $T_{\textrm{c}}$  under pressure up to about 4.6 GPa \cite{BERTHIER1976SolidStateComm_NbSe2_Pressure}. Suderow {\it et al.} proposed a peculiar interplay among superconductivity, CDW order, and Fermi surface complexity, based on the mismatch between the suppression of $T_{\textrm{CDW}}$ at 5 GPa and the maximum $T_{\textrm{c}}$ at 10.5 GPa \cite{Suderow2005PRL_NbSe2_Pressure_FS}. Chatterjee {\it et al.} \cite{Chatterjee2015} studied the effect of transition metal doping on the CDW state using ARPES, X-ray diffraction, STM tunneling, and resistivity, and showed that short range CDW order and an energy gap remained at high temperatures and high disorder beyond the phase coherence transition.

Another way to probe CDW and SC states 
is to introduce non-magnetic point-like scatterers\cite{Mutka1983PRB_NbSe2_TaS2_TaSe2_e-irr, Grest1982PRB_Tc_increase_with_CDW, Hirschfeld1993PRB, WangHirschfeldMishra2013PRB, Cho2016ScienceAdvances_BaK122_e-irr}. Electron irradiation, which has been shown to create pure atomic disorder without doping the system as deduced from Hall effect measurements,
is an effective tool to influence both the superconductivity and other orders \cite{Prozorov2014PRX,Prozorov2017Tc-TN, Cho2018SST_review_e-irr}. Moreover, independent measurements of $T_{\textrm{c}}$, $T_{\textrm{CDW}}$ and low-temperature London penetration depth in samples with controlled disorder become powerful techniques that can distinguish different types of superconducting pairing such as $d-$wave, $s_{\pm}$, and $s_{++}$ pairing \cite{Hirschfeld1993PRB, EfremovHirschfeld2011PRB, WangHirschfeldMishra2013PRB}. Indeed, this approach was successfully used in various iron-based superconductors \cite{Prozorov2014PRX, Mizukami2014NatureComm, StrehlowProzorov2014PRB, Cho2016ScienceAdvances_BaK122_e-irr}. According to early studies of the effect of electron irradiation on NbSe$_2$ by Mutka {\it et al.} \cite{Mutka1983PRB_NbSe2_TaS2_TaSe2_e-irr}, an increase of $T_{\textrm{c}}$ was reported but attributed to inhomogeneous superconductivity. This result was theoretically discussed by  Grest {\it et al.} \cite{Grest1982PRB_Tc_increase_with_CDW} and Psaltakis {\it et al.}, \cite{Psaltakis1984SSP_TC_increase_with_CDW}, but  direct evidence  determining the effect of homogeneously distributed disorder on the interplay between the CDW and SC states is still missing.

In this article, we systematically investigate the effect of controlled point-like disorder on superconductivity and CDW order in 2$H$-NbSe$_2$. The disorder is generated by applying 2.5 MeV electron irradiation with different doses. For each dose, the changes in $T_{\textrm{c}}$, residual resistivity, Hall coefficient, and London penetration depth are measured.  For low irradiation doses,  $T_{\textrm{c}}$ shows nonmonotonic behavior, first increasing slightly and then decreasing until a critical dose of 1.0 C cm$^{-2}$ where it drops abruptly.  At this critical dose,  the long-range CDW feature in resistivity disappears  as well.  The vanishing of $T_{\textrm{CDW}}$  appears to be discontinuous. Upon further irradiation, we find the existence of persistent short-range CDW correlations based on X-ray scattering and Hall resistivity measurements, and  attribute the abrupt drop in $T_{\textrm{c}}$ to the loss of coherence of the phase-coherent CDW order. Among various possible mechanisms, we conclude that the effect of the reconstruction of the electronic structure by the CDW leads to a rapid change of electron-phonon scattering at the critical dose of 1.0 C cm$^{-2}$, explaining a remarkable qualitative change in the Hall effect and an abrupt drop of $T_{\textrm{c}}$. This represents a clear evidence for a special role of the coherent  CDW state coupling to superconductivity. Furthermore, the change in $T_{\textrm{c}}$ provides important information on the nature of the pairing both within and outside of the long-range CDW state. Upon irradiation above  the critical dose, $T_{\textrm{c}}$ continuously decreases down to the largest dose applied, suggesting a substantial degree of gap anisotropy. The low temperature London penetration depths of three post-irradiated  samples consistently show exponentially saturating behavior below 0.2 $T_{\textrm{c}}$, with gaps that increase with disorder and are therefore consistent with this picture.

\begin{figure}[htb]
\includegraphics[width=8.5cm]{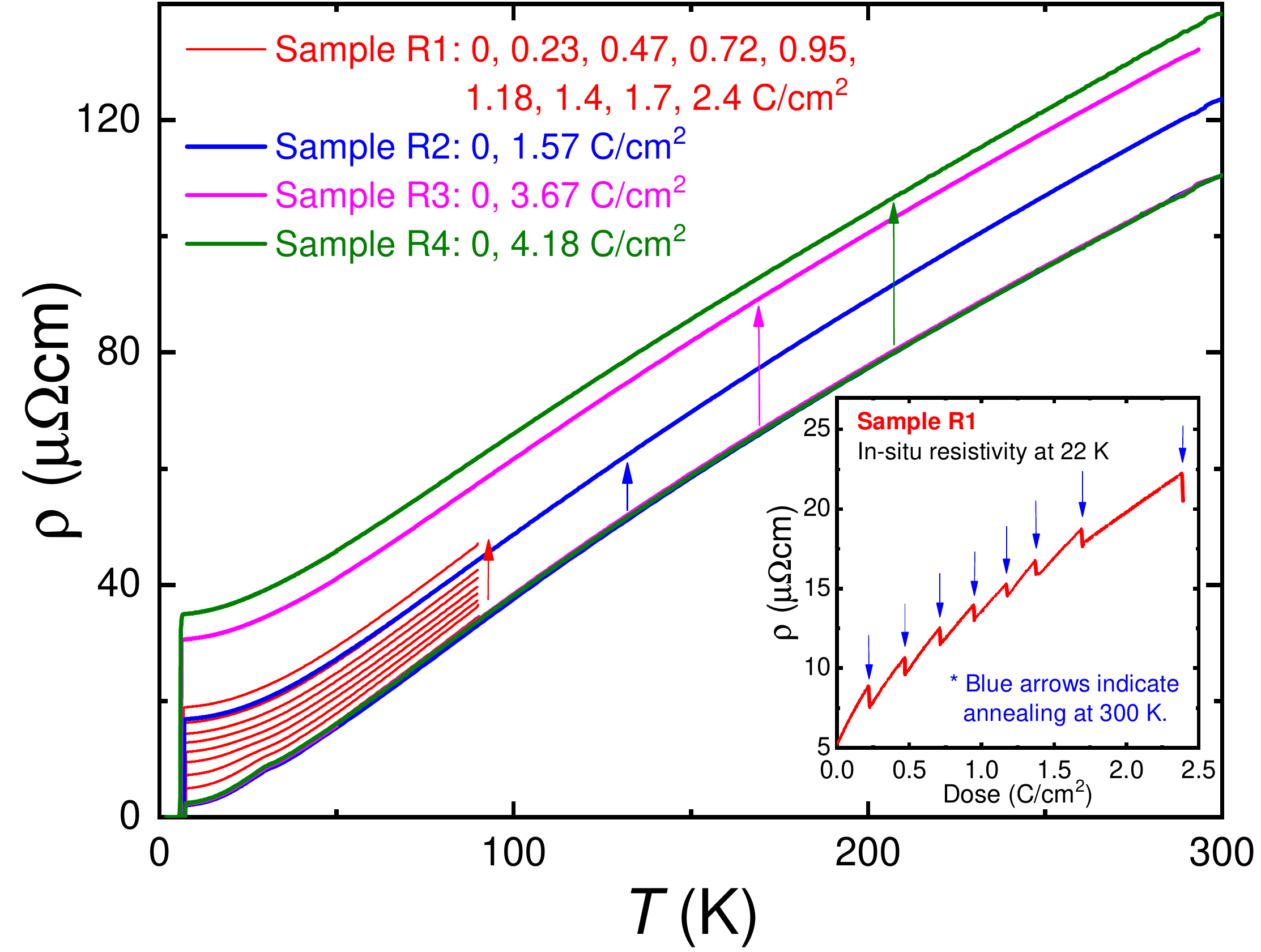}
\caption{\textbf{Temperature dependence of resistivity upon electron irradiation.} Resistivities of four different samples: R1 (0, 0.23, 0.47, 0.72, 0.95, 1.18, 1.4, 1.7, and 2.4 C cm$^{-2}$), R2 (0, 1.57 C cm$^{-2}$), R3 (0, 3.67 C cm$^{-2}$), and R4 (0, 4.18 C cm$^{-2}$).  Note that all 0-dose curves for samples R1, R2, R3, and R4 are coincident. Overall resistivity increase with increasing dose was consistently seen for all samples, as shown by the arrows. The inset shows in-situ measurement of resistivity of sample R1 as a function of dose during electron irradiation at 22 K. The blue arrows indicate stops in irradiation, during which the sample was extracted from irradiation chamber and characterized. Partial annealing of about 30 to 40$\%$ of resistivity increase occurred on warming the sample to room-temperature and subsequent cool-down to 22 K.}
\label{fig1}
\end{figure}

\section{Results}

\subsection{Effect of electron irradiation on resistivities}
\begin{figure}[htb]
\includegraphics[width=8.5cm]{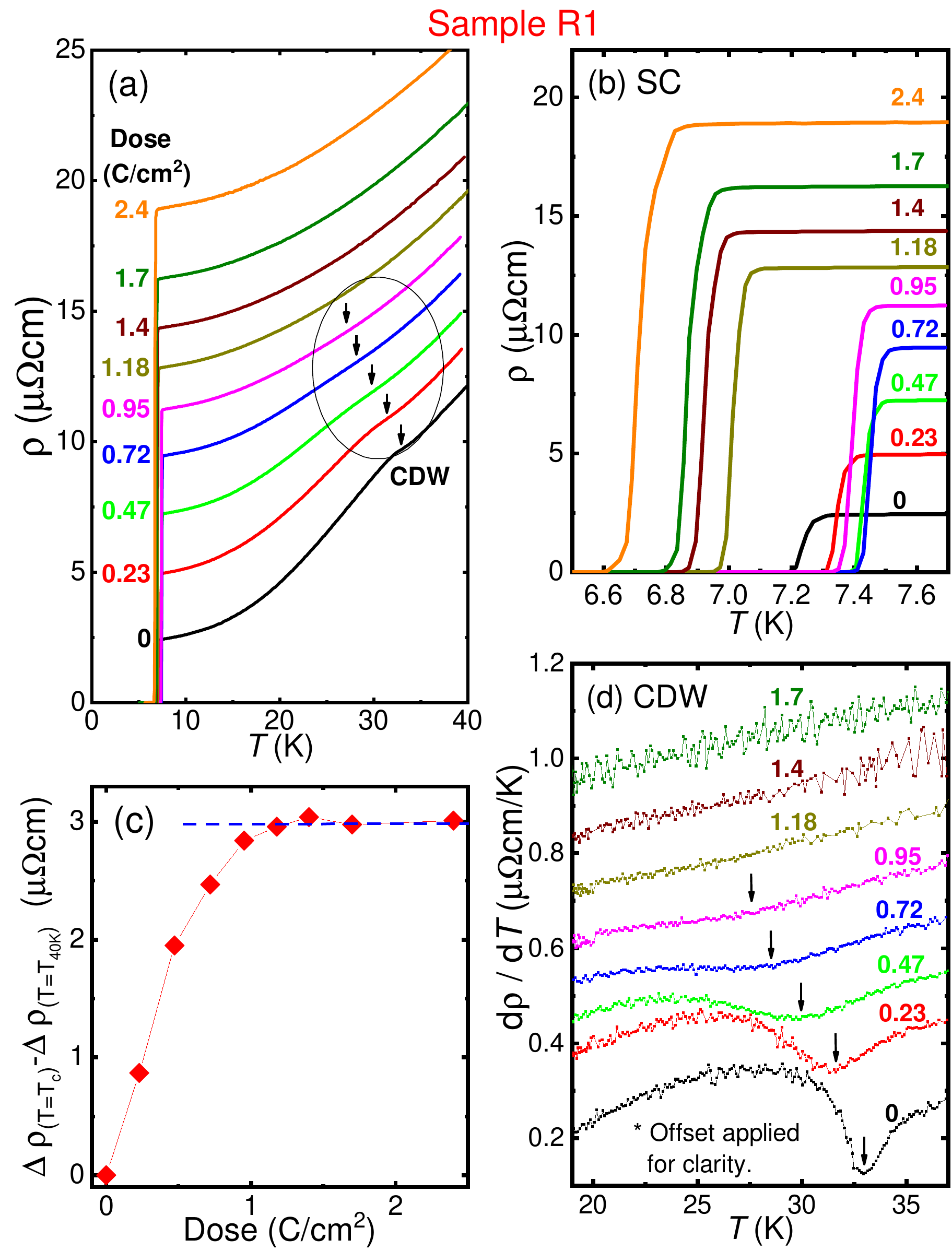}
\caption{\textbf{Resistivity measurement conducted on sample R1.} (a) The temperature dependence of resistivity upon irradiation. Overall, the resistivity above the CDW transition was parallel-shifted upward suggesting that preservation of Matthiessen's rule only occurs above the transition (see text for further discussion). (b) The zoom of superconducting transition area that shows the increase and subsequent decrease of $T_{\textrm{c}}$ upon irradiation. (c) $\Delta \rho_{(T = 7.6 K)} - \Delta \rho_{(T = 40 K)}$ that shows diappearance of the CDW above 1.0 C cm$^{-2}$. (d) The derivative of resistivity with respect to temperature that manifests the location of the CDW transitions.}
\label{fig2}
\end{figure}

Electron irradiation (maximum dose of 8.93 C cm$^{-2}$) effectively introduces artificial disorder into the system, resulting in the substantial increase of residual resistivity in the normal state, as shown in Fig.~\ref{fig1}. Above 40 K without long-range CDW order, the increase of resistivity is rather constant. However, near and below 40 K, a violation of Matthiesen's rule was observed due to the presence of the CDW phase. For  cases with high doses of irradiation ($>$ 1.0 C cm$^{-2}$) where the CDW feature in resistivity was completely suppressed due to disorder,  Matthiesen's rule was obeyed over the entire temperature region of the normal state. To investigate how effectively the electron irradiation introduces defects, the in-situ resistivity of sample R1 was measured during the irradiation at 22 K (inset of Fig.~\ref{fig1}). It increases monotonically with increasing dose of irradiation. The blue arrows indicate when the irradiation stopped and room temperature annealing occurred. About 30 - 40 $\%$ annealing occurred for each case. For each dose (blue arrow), the sample was removed from the irradiation chamber and moved to a different cryostat for measurement of the temperature dependent resistivity as shown in Fig.~\ref{fig1} and \ref{fig2} (b).

With increasing irradiation dose, both the superconducting and CDW phases were substantially affected. As shown in Fig.~\ref{fig2} (a), $T_{\textrm{CDW}}$ (kink feature marked by arrow) decreases with increasing irradiation and disappears after 1.0 C cm$^{-2}$. The behavior of the CDW feature is more clearly shown in a plot of $d \rho / dT$ versus temperature (Fig.~\ref{fig2} (d)). The important fact is that the feature associated with the CDW transition disappears at finite temperature of 27 K instead of continuing down to zero Kelvin. This result suggests the absence of a quantum critical point with disorder, in contrast to the previous pressure study by Feng {\it et al.}. \cite{Feng2012PNAS_CDW_QCP}. Fig.~\ref{fig2} (b) is an enlargement of the low-temperature part of Fig.~\ref{fig2} (a) that  shows the change of $T_{\textrm{c}}$. It is clearly seen that $T_{\textrm{c}}$ initially increases and then decreases upon irradiation. All the values of $T_{\textrm{c}}$ and $T_{\textrm{CDW}}$ for sample R1 are summarized in Fig.~\ref{fig4} along with $T_{\textrm{c}}$'s from other samples (R2, R3, R4, P1, P2, P3). The $x$-axis of Fig.~\ref{fig4} is the increase of resistivity at 40 K, $\Delta \rho _{(T = 40 K)}$, upon irradiation (representing increased disorder). With increasing dose of irradiation up to 1.0 C cm$^{-2}$ ($\Delta \rho _{(T = 40 K)}$ = 7.3 $\mu \Omega$cm), $T_{\textrm{c}}$ gradually increases from 7.25 K to 7.45 K, and then starts decreasing back to 7.3 K while $T_{\textrm{CDW}}$ monotonically decreases. Upon further irradiation, the feature associated with the CDW transition disappears and simultaneously $T_{\textrm{c}}$ abruptly drops by 0.3 K, indicating strong correlation between the superconducting and CDW phases. Note the mismatch between the maximum $T_{\textrm{c}}$ and the disappearance of the CDW feature, suggesting a complex interplay between the two phases potentially including other factors such as complicated Fermi surfaces. Upon further irradiation, $T_{\textrm{c}}$ continues to decrease toward about 33$\%$ of its pristine value for the maximum electron dose of 8.93 C cm$^{-2}$ ($\Delta \rho _{(T = 40 K)}$ = 61 $\mu \Omega$cm) as shown in the full phase diagram in Supplementary Figure~1.

\begin{figure}[htb]
\includegraphics[width=8.5cm]{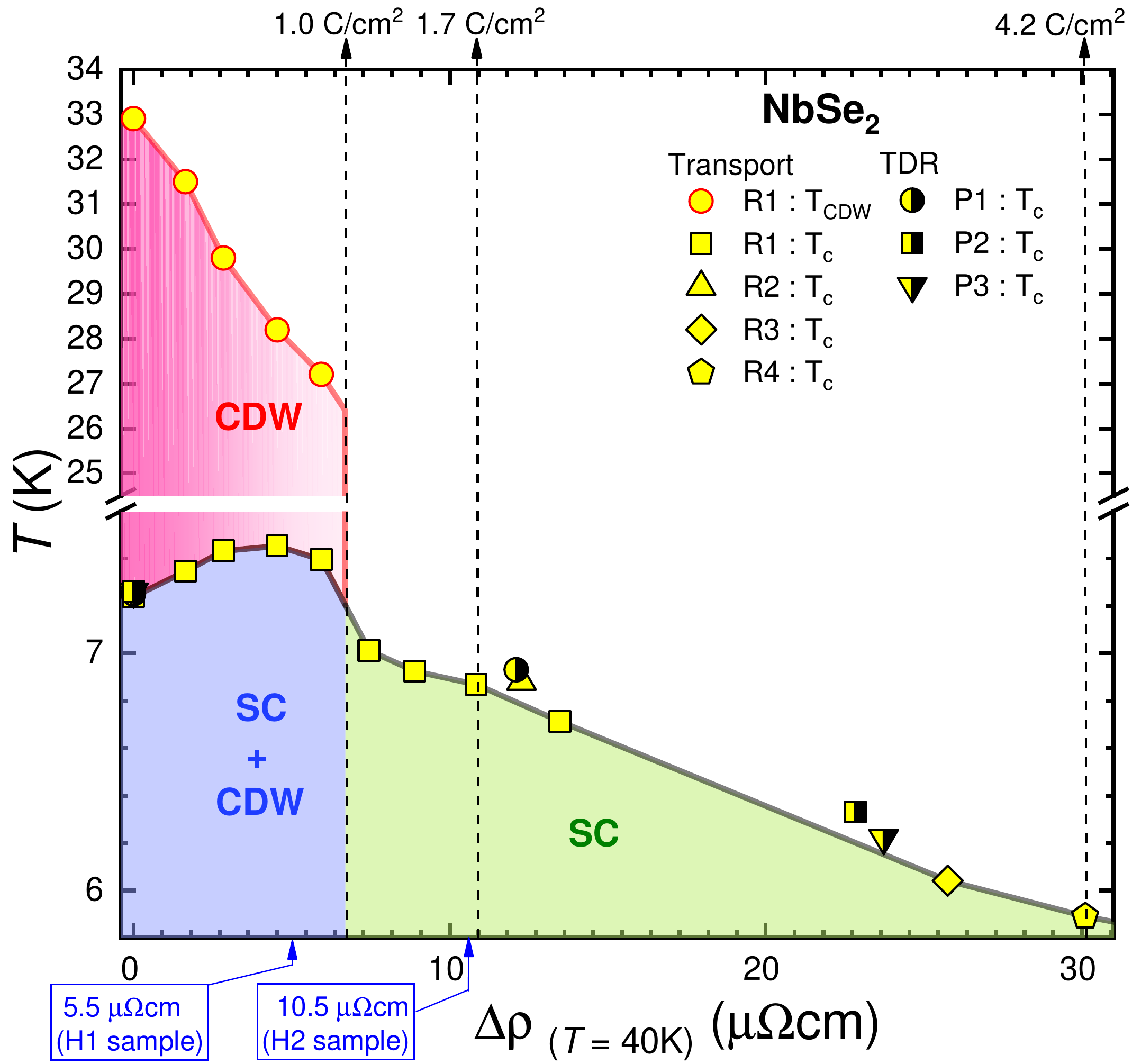}
\caption{\textbf{Temperature-dependent penetration depth changes $\Delta\lambda$ upon electron irradiations.} Results of three samples before and after irradiation: P1 (0 and 1.77 C cm$^{-2}$), P2 (0 and 3.47 C cm$^{-2}$), and P3 (0, 3.34, and 8.93 C cm$^{-2}$). (a) Wide temperature span of $\Delta\lambda$ that shows a substantial decrease of $T_{\textrm{c}}$ upon irradiation. (b-d) The low temperature part of $\Delta\lambda$ of (b) P1, (c) P2, and (d) P3 samples before and after irradition. All data clearly show the saturating behavior below 0.2 $T/T_{\textrm{c}}$, supporting the presence of s-wave type superconducting gaps.}
\label{fig3}
\end{figure}

\subsection{Effect of electron irradiation on the London penetration depth}
\begin{figure}[htb]
\includegraphics[width=8.5cm]{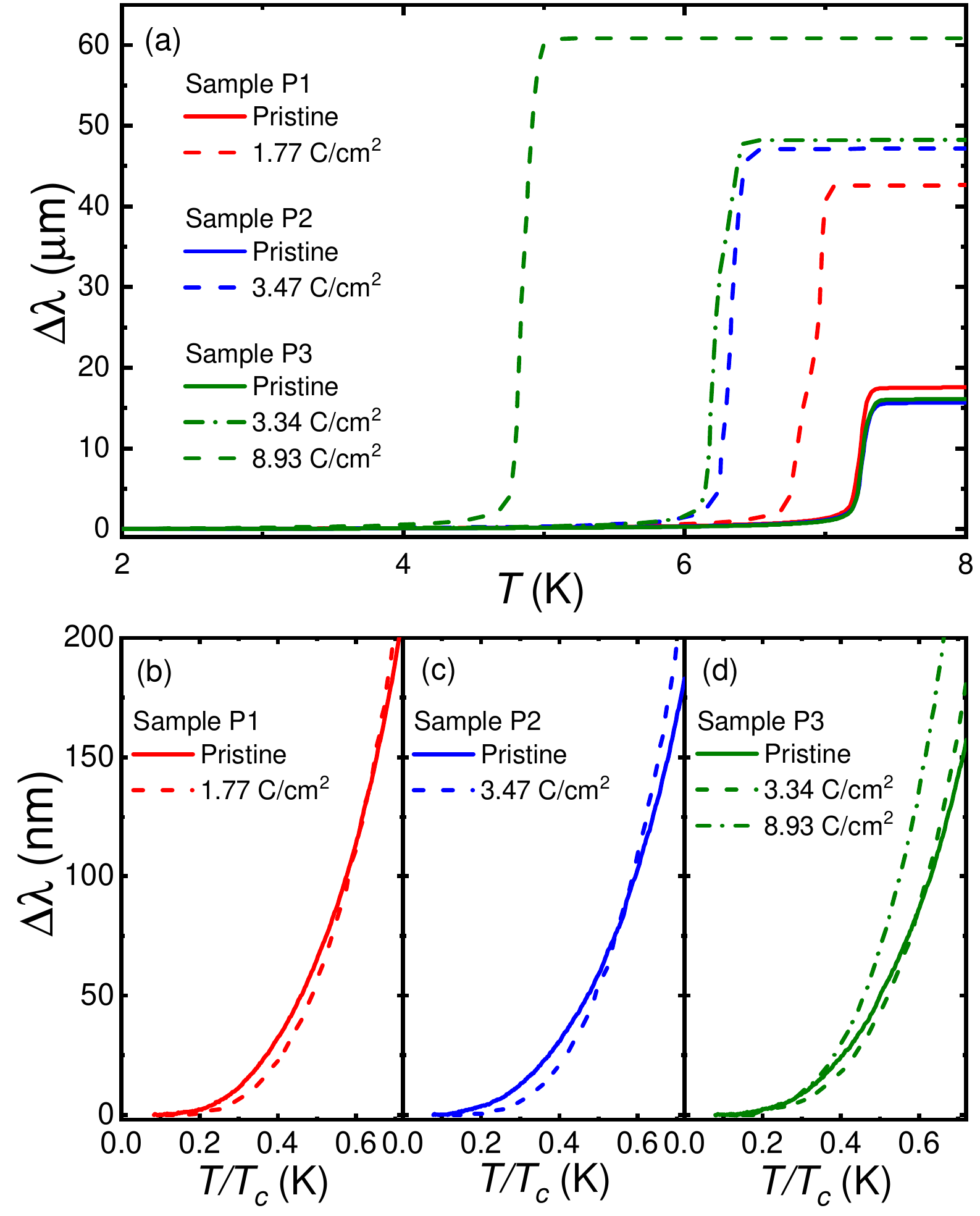}
\caption{\textbf{Temperature versus $\Delta \rho_{(T = 40 K)}$ phase diagram upon electron irradiations.} For low doses of irradiation up to 1.0 C cm$^{-2}$, $T_{\textrm{c}}$ varies nonmonotonically, while $T_{\textrm{CDW}}$ monotonically decreases. When the resistivity feature of CDW disappears around 1.0 C cm$^{-2}$ (shown in Fig.~\ref{fig2} (b) and (d)), $T_{\textrm{c}}$ suddenly drops by 0.3 K, indicating the strong correlation between superconductivity (SC) and CDW order. Upon further irradiation, $T_{\textrm{c}}$ monotonically decreases. The full phase diagram up to the highest dose of 8.93 C cm$^{-2}$ is shown in Supplementary Figure~1. Hall resistivity is measured in two samples H1 and H2 (Fig.~\ref{fig6} (a) and (b)). Two blue arrows in $x$-axis indicate their locations in the phase diagram, based on the increase in resistivity (Fig.~\ref{fig6} (c) and (d)).}
\label{fig4}
\end{figure}

Fig.~\ref{fig3} exhibits the temperature dependence of the London penetration depth ($\Delta \lambda$) of P1, P2 and P3 samples upon irradiation. The low temperature saturation is clearly seen below 0.2 $T$/$T_{\textrm{c}}$ for all samples before and after irradiation, suggesting the presence of s-wave type superconducting gaps. Interestingly, the saturation tendency gets stronger  after irradiation, suggesting that the initial anisotropic gaps get more isotropic due to the gap-smearing effect of point-like disorder. Note that the irradiation doses shown correspond to residual resistivities $\Delta \rho$ beyond the initial enhancement of $T_{\textrm{c}}$ due to competition with the CDW phase, so that a uniform enhancement of $\Delta$ is not the main cause of the saturation in $\Delta\lambda$. In addition, a substantial decrease of $T_{\textrm{c}}$ from 7.25 K to 4.8 K (about 33$\%$ decrease) was found in sample P3 upon 8.93 C cm$^{-2}$. All the $T_{\textrm{c}}$ suppressions from these three samples (P1, P2, P3) are summarized in Fig.~\ref{fig4}. Since we cannot directly obtain $\rho_{(T = 40 K)}$ for P1, P2, and P3, we used the relation between dose and $\Delta \rho$ obtained from transport-measured samples (R1, R2, R3, and R4)  as shown in Supplementary Figure~2 (a). The substantial decrease of $T_{\textrm{c}}$ and exponential-like saturation of $\Delta \lambda$ can be explained with multiband s-wave type superconducting gap with some amount of interband coupling.

\subsection{Phase diagram upon electron irradiation}
\begin{figure}[htb]
\includegraphics[width=8.5cm]{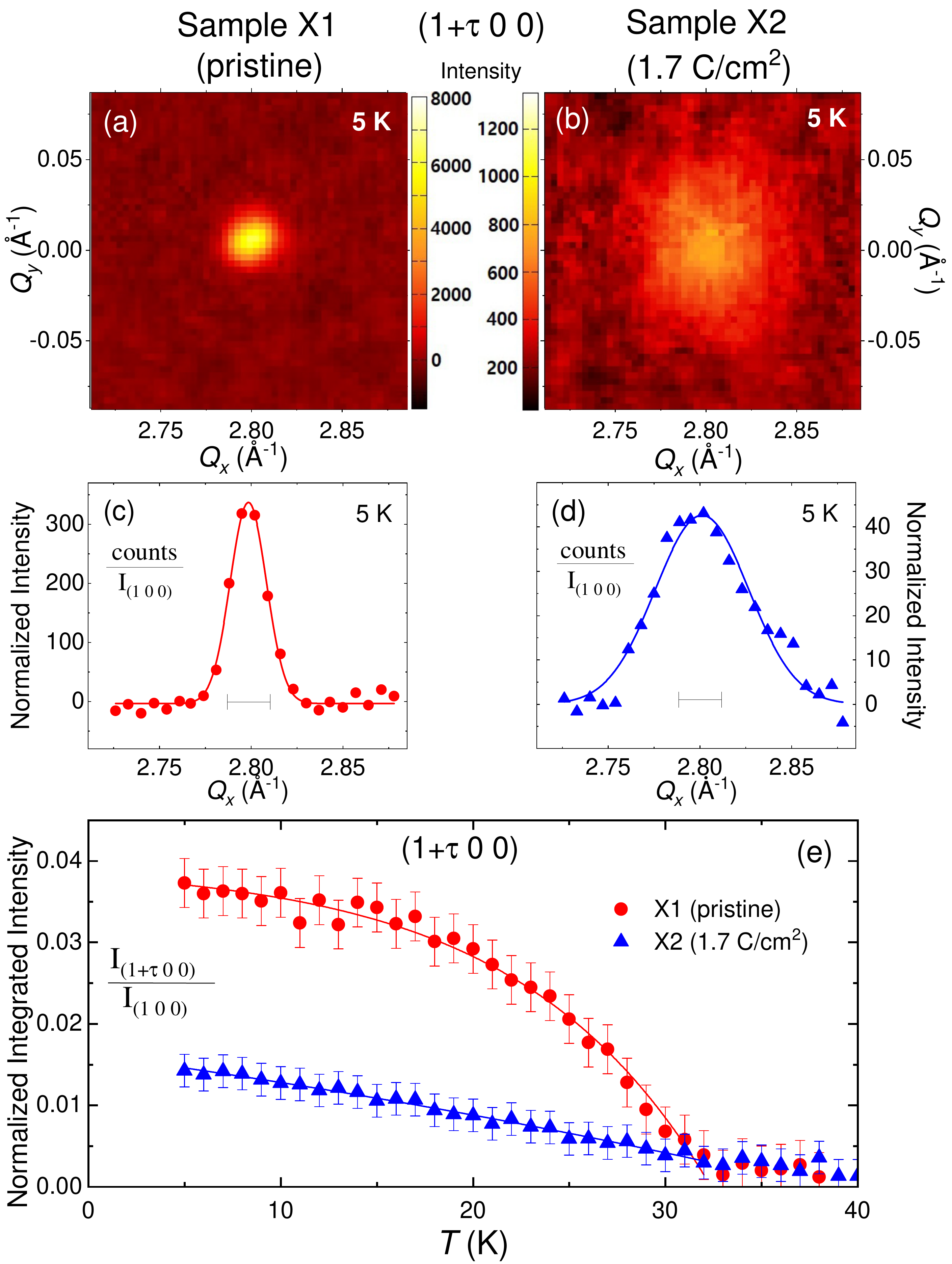}
\caption{\textbf{High-energy x-ray diffraction measurement of the CDW Bragg peak (1+$\tau$ 0 0) with $\tau$ = 1/3.} Results of two NbSe$_2$ samples (X1: pristine, X2: 1.7 C cm$^{-2}$). (a-b) Diffraction patterns of X1 and X2 recorded by the 2-dimensional detector in the ($H$ $K$ 0) plane with intensity encoded in a linear colour scale for each detector pixel. (c-d) Cuts along the longitudinal direction $Q_x$, integrated along the transverse direction $Q_y$ and normalized to the integrated intensity I$_{(1 0 0)}$ of the weak (1 0 0) Bragg peak of the regular chemical lattice which is three orders of magnitude less intense than the strong (1 1 0) Bragg peak. The bars represent the instrumental resolution full-width half-maximum determined from the (1 0 0) Bragg peak. (e) Temperature dependence of the normalized integrated intensity of the CDW Bragg peak.}
\label{fig5}
\end{figure}

\begin{figure}[htb]
\includegraphics[width=8.5cm]{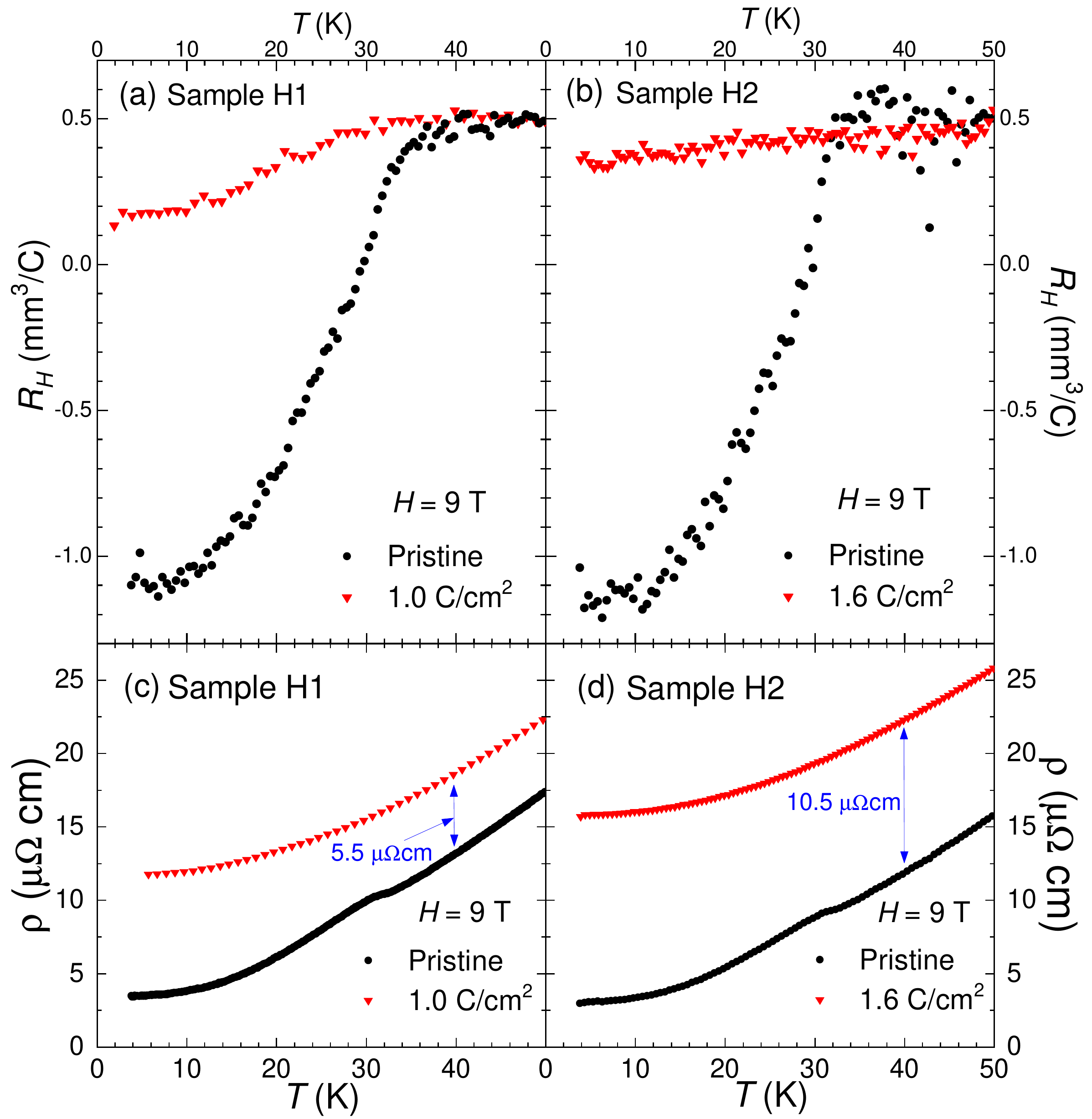}
\caption{\textbf{Temperature dependence of Hall resistivities of two NbSe$_2$ samples measured under 9 T.} (a) Sample H1  before and after 1.0 C cm$^{-2}$ irradiation and (b) Sample H2 before and after 1.6 C cm$^{-2}$ irradiation. (c-d) In-field Resistivities were also measured for both samples that clearly show the increase of disorder: $\Delta \rho_{(T = 40 K)}$ = 5.5 and 10.5 $\mu \Omega$cm for samples H1 and H2, respectively.}
\label{fig6}
\end{figure}

 Fig.~\ref{fig4} shows the temperature versus $\Delta \rho_{(T = 40 K)}$ phase diagram of superconductivity and charge density wave upon electron irradiation obtained from seven samples. Upon initial irradiation up to 1.0 C cm$^{-2}$, an anticorrelation of $T_{\textrm{CDW}}$ and $T_{\textrm{c}}$ was observed, which is most naturally interpreted in terms of strong competition between the superconducting and CDW phases. However, after $T_{\textrm{c}}$ reaches its maximum, both $T_{\textrm{CDW}}$ and $T_{\textrm{c}}$ decrease until the CDW phase abruptly disappears at a critical irradiation dose of 1.0 C cm$^{-2}$, where $T_{\textrm{c}}$ also drops discontinuously. The simplest explanation of the nonmonotonic behavior of $T_{\textrm{c}}$ in the CDW coexistence phase is that the initial increase is due to the competition between the superconductivity and CDW phases. The effect of disorder on this competition was studied already by Grest {\it et al.} \cite{Grest1982PRB_Tc_increase_with_CDW} and Psaltakis {\it et al.} \cite{Psaltakis1984SSP_TC_increase_with_CDW}.  Within this weak-coupling approach, non-magnetic disorder suppresses the CDW rapidly, and since the CDW order is competing for Fermi surface with superconductivity, $T_{\textrm{c}}$ increases.  Note that these theoretical calculations assumed an isotropic $s$-wave gap; within their approximations, the superconducting $T_{\textrm{c}}$ would have saturated when CDW order vanished, due to Anderson's theorem. However, it is clear from Fig.~\ref{fig4} that disorder continues to suppress $T_{\textrm{c}}$ long after the CDW order is gone; this implies that the $s$-wave gaps have quite different amplitudes (and, possibly, anisotropy) and substantial interband pairing. Furthermore, as will be shown in Figs.~\ref{fig5} and \ref{fig6}, we found from the Hall resistivity and X-ray scattering that the short-range CDW phase still survives long after the critical dose of 1.0 C cm$^{-2}$.

\subsection{X-ray diffraction upon electron irradiation}
Figure~\ref{fig5} shows the characterization of the CDW of samples X1 (pristine) and X2 (1.7 C cm$^{-2}$) by high-energy x-ray diffraction. The structure of NbSe$_2$ consists of layers of Nb atoms surrounded by 6 Se atoms and the Nb atoms located in the corners of the hexagonal unit cell \cite{MAREZIO1972JSolidStatChem}.  The CDW  displaces the six nearest Nb neighbors of every third Nb atom, yielding a superstructure with tripling of the unit-cell dimensions in both a and b directions, and corresponding propagation vectors ($\tau$ 0 0) and (0 $\tau$  0) with $\tau$ = 1/3~\cite{Malliakas2013JACS_NbSe2_CDW}. For both samples, the CDW Bragg peaks are observed in all measured Brillouin zones at low temperature. The CDW Bragg peaks are resolution limited, as illustrated in Figs.~\ref{fig5} (a) and (c) for sample X1, whereas they show a significant broadening in panels (b) and (d) for sample X2 which is temperature-independent. Note that the intensity scales in panels (a) and (c) are about seven times larger than those in panel (b) and (d). From the peak widths determined by Gaussian fits to the cuts shown in Figs.~\ref{fig5} (c) and (d), the correlation length of the CDW is estimated to be about 80~\AA~for the irradiated sample X2 and a lower limit of 200~\AA~for the pristine sample X1. The temperature dependence of the normalized integrated CDW Bragg peak intensity shown in Fig.~\ref{fig5} (e) represents the square of the CDW order parameter and is clearly consistent with a second-order phase transition at $T_{\textrm{CDW}}$ = 33 K and long-range order for sample X1. In contrast, the CDW intensity of the irradiated sample X2 increases continuously with decreasing temperature without a clear onset, indicating a cross-over like behavior. Together with the reduced correlation length, it is clear that the CDW manifests only short-range order in the irradiated sample X2, although the strength of the CDW is almost similar to the pristine sample X1 with the integrated Bragg peak intensity only reduced by 65~\%~at low temperature.

The CDW appears in the same manner with comparable strength in irradiated samples but with a reduced correlation length. The irradiation induced defects likely form barriers or pinning centers for boundaries of the CDW state and prevent a coherently ordered state beyond these defects when the CDW state develops with decreasing temperature. The crossover-like temperature dependence of the short-range CDW without a clear onset observed for the sample X2 is consistent with the lack of a well-defined feature or signature of the CDW in transport measurements for samples with radiation levels above the critical dose of 1.0 C cm$^{-2}$.

\subsection{Hall resistivities upon electron irradiation}
Hall resistivities were measured for samples H1 and H2 as shown in Fig.~\ref{fig6} (a) and (b). For sample H1, two measurements were conducted before and after irradiation with 1.0 C cm$^{-2}$, and for sample H2, before and after irradiation with 1.6 C cm$^{-2}$. First of all, the Hall resistivity of both pristine samples shows a sign-change below $T_{\textrm{CDW}}$ = 33 K, consistent with previous reports \cite{YAMAYA1972SolidStatComm_NbSe2_Hall,Bel2003PRL_Nernst_NbSe2}. This indicates an increase of mobility in the CDW phase, consistent with previous reports in resistivity and Nernst effect \cite{Bel2003PRL_Nernst_NbSe2},  due to an opening of the pseudogap \cite{Evtushinsky2008PRL_TaSe2_NbS2_APRES_Hall}. The change of in-plane resistivity, $\rho_{xx}$, as shown in  in panels (c) and (d), was used to accurately calibrate the amount of disorder, yielding $\Delta \rho_{(T = 40 K)}$ = 5.5 $\mu \Omega$cm for sample H1 and 10.5 $\mu \Omega$cm for H2. These values allow us to place the samples before and after the critical transition, respectively, as shown in the phase diagram (blue arrows in $x$-axis) of Fig.~\ref{fig4}. The CDW transition is clearly seen for sample H1 in Fig.~\ref{fig6} (a), consistent with the observation of a feature in the resistivity derivative. For sample H2, however, the feature at the CDW transition almost disappears in Fig.~\ref{fig6} (b) although a slight slope change can be noticed at 30 K. This is consistent with the disappearance of the long - range CDW feature (resistivity) and the presence of a short -range CDW  (X-ray scattering) above the critical dose. Another important fact is that the Hall resistivity above 40 K did not change upon irradiation. This implies that the defects introduced by electron irradiation do not change the electronic carrier density above 40 K, but only increase the scattering rate.

\section{Discussion}
Evidence for an anisotropic superconducting gap in NbSe$_2$ is provided, e.g. by STM measurements, which show a significantly broadened gap edge \cite{GuillamonSuderow2008PRB_NbSe2_STM_modulation}. In addition, the flattening of the low-$T$ penetration depth upon irradiation at doses corresponding to the pure superconducting phase is evident in Fig.~\ref{fig3}. Since the smallest gap in the system will determine the asymptotic low-$T$ exponential dependence, it suggests that disorder is increasing the minimum gap, i.e. gap averaging. The nonmonotonic behavior of $T_{\textrm{c}}$ in the CDW/SC coexistence phase can therefore be understood simply by assuming that the effect of losing competition from the CDW is overcome by the gap averaging effect before the CDW disappears. It should be noted, however, that the behavior with pressure is also nonmonotonic \cite{Suderow2005PRL_NbSe2_Pressure_FS}  In this case the reason for the continued suppression of $T_{\textrm{c}}$ is less obvious, and the pressure dependence of the couplings of various phonons may be necessary to explain the complete behavior quantitatively.

 The possibility of a first-order transition at the disappearance of CDW order is also intriguing and recalls the question of the CDW mechanism. A simple Fermi surface nesting model \cite{Wilson1977} fails to explain the CDW ordering vector, and is therefore not appropriate for NbSe$_2$\cite{Valla2004,Weber2011PRL_NbSe2_CDW_Phonon}.   Similarly, a saddle point-driven CDW instability proposed by Rice and Scott \cite{Rice1975} has been ruled out by ARPES \cite{Rossnagel2001}. However, a generalized Fermi surface nesting model, which includes the strong anisotropy in the electron-phonon matrix elements, does capture the correct CDW ordering vector \cite{Doran1978, Flicker2015NatComm_NbSe2_CDW}. The generalized Fermi-surface nesting model is still effectively a weak coupling model, where a strong momentum dependence of the electron-phonon matrix elements modifies the peak position of the charge susceptibility. Hence, from a weak coupling perspective, a disorder-driven first order transition, as apparently observed here, appears to be a natural one. This is because the CDW is a Stoner-type instability, where with increasing disorder the charge susceptibility at ordering vectors should drop below a critical value corresponding to ordering.

Observation of a quantum critical point under pressure might be taken as evidence against the idea of first order transition\cite{Feng2012PNAS_CDW_QCP}. However, one should keep in mind that pressure also changes the bare electronic structure, which does not happen in case of point like impurities. Disorder  is often thought to drive a first order transition,  e.g. in the manganites \cite{Dagotto2004}. We note that Chatterjee {\it et al.} \cite{Chatterjee2015} deduced a smooth decay of CDW order with chemical substitution, but in fact their data are entirely consistent with ours because of the relatively small number of doping levels  studied in that work.

We cannot definitively rule out the possibility that the feature observed in transport, here identified as the signature of long-range CDW order, simply becomes too weak to observe because of broadening due to significant short-range fluctuations, as observed in Ref. \onlinecite{Chatterjee2015}.  However, our new observation of a concomitant abrupt drop in $T_{\textrm{c}}$ suggests that a thermodynamic transition is indeed taking place at  this critical value of disorder. Unlike incommensurate CDW systems, commensurate CDW transitions as in NbSe$_2$ in the presence of quenched disorder may occur \cite{Nie2014}.  The ordered phase in such a situation breaks translational symmetry discretely, so that a second order transition with exponents dependent on the order of the commensurability is allowed, but this can be preempted by a first order transition, as apparently observed here.

We investigated the interplay between the  CDW and superconducting phases in 2$H$-NbSe$_2$ by measuring the resistivity and penetration depth before and after electron irradiation. Upon initial irradiation, $T_{\textrm{c}}$ increased from 7.25 K to 7.45 K, and then decreased while $T_{\textrm{CDW}}$ monotonically decreased. This indicates a complex interplay between two phases with potential other factor such as a complicated Fermi surface. Upon further irradiation, the feature associated with the  CDW transition disappeared at finite temperature. When the CDW feature disappears, $T_{\textrm{c}}$ abruptly dropped by 0.2 K, indicating strong correlation between two phases and suggesting a first order, disorder driven phase transition. Further irradiation up to 8.93 C cm$^{-2}$ effectively and monotonically decreased $T_{\textrm{c}}$ down to 4.8 K (about 33$\%$ of its pristine value), suggestive of the averaging of an anisotropic $s$-wave superconducting order parameter. According to X- ray scattering and Hall resistivity studies, the short-range CDW is still present after the critical dose of $\sim$ 1.0 C cm$^{-2}$ ($\sim$ 7.3 $\mu \Omega$cm) indicating that the effect of electron irradiation decreases the coherence of the CDW phase. The low-temperature penetration depth shows exponential-like behavior below 0.2 $T$/$T_{\textrm{c}}$ for all samples before and after irradiation. The combined results of resistivity and penetration depth can be explained with multiband anisotropic s-wave type superconducting gaps with some amount of interband coupling.

\section{Methods}
\subsection{Crystal growth}
The single crystals of 2$H$-NbSe$_2$ from Bell Laboratories were grown using the usual iodine vapor transport technique and are known to be of high quality (RRR $\sim$ 40). These are the samples from the same batch as used in Ref. \cite{FletcherProzorov2007PRL_NbSe2_TDR}. Four-probe measurements of in-plane resistivity were performed for four samples (R1, R2, R3, and R4). Samples for resistivity measurements had dimensions of (1-2) $\times$ 0.5 $\times$ (0.02-0.1) mm$^3$. Electrical contacts to samples prior to irradiation were made by soldering 50-$\mu$m silver wires with indium and mechanically strengthened by silver paste as described elsewhere \cite{Tanatar2016}. For in-situ resistivity measurement during the electron-irradiation at 22 K, R1 sample was mounted on a Kyocera chip as shown in the inset of Fig.~\ref{fig1} and measured during irradiation.

\subsection{Resistivity measurements}
Simultaneous Hall effect and resistivity measurements were performed on samples H1 and H2 mounted in 5-probe configuration using the same contact making technique as in resistivity measurements. Measurements were taken in {\it Quantum Design} PPMS in constant magnetic fields +9T and -9T. Same samples with the same contacts were measured before and after irradiation, thus excluding geometric factor errors in quantitative comparison.

\subsection{London penetration depth}
The in-plane London penetration depth $\Delta \lambda$ (T) of three other samples (P1, P2, and P3) was measured before and after irradiation using a self-oscillating tunnel-diode resonator technique \cite{Prozorov2000PRB, Prozorov2006SST, ProzorovKogan2011RPP}. The samples had typical dimensions of 0.5  $\times$ 0.5  $\times$ 0.03 mm$^3$.

\subsection{X-ray diffraction}
The high-energy x-ray diffraction study was performed at station 6-ID-D at the Advanced Photon Source, Argonne National Laboratory. The use of x-rays with an energy of 100.5 keV minimizes sample absorption and allows to probe the entire bulk of the sample using an incident beam with a size of 0.5 $\times$ 0.5 mm$^2$, over-illuminating the sample. The samples were held on Kapton tape in a Helium closed-cycle refrigerator and Helium exchange gas was used. Extended regions of selected reciprocal lattice planes were recorded by a MAR345 image plate system positioned 1468 mm behind the sample as the sample was rocked through two independent angles up to $\pm 3.2^\circ$ about axes perpendicular to the incident beam \cite{KreyssigGoldman2007PRB_x-ray_Ce2Fe17}.

\subsection{Electron irradiation}
The 2.5 MeV electron irradiation was performed at the SIRIUS Pelletron facility of the Laboratoire des Solides Irradies (LSI) at the Ecole Polytechnique in Palaiseau, France
\cite{VanDerBeekKonczykowski2013JPCS_e-irr}. The acquired irradiation dose is conveniently measured in C cm$^{-2}$, where 1 C cm$^{-2}$ = 6.24 $\times$ 10$^{18}$ electrons/cm$^2$.

\subsection{Data availability}
The authors declare that all data supporting the findings of this
study are available within the article and its supplementary information files or
from the corresponding author upon reasonable request.

\bibliographystyle{apsrev4-1}

\section*{Acknowledgements}
The authors are grateful to H. Suderow, S. Raghu, S. Sachdev, and R. J. Cava for useful discussions.  The work in Ames Laboratory was supported by the U.S. Department of Energy (DOE), Office of Science, Basic Energy Sciences, Materials Science and Engineering Division. Ames Laboratory is operated for the U.S. DOE by Iowa State University under contract DE-AC02-07CH11358. P.J.H. was supported by NSF Grant No. DMR-1407502. We thank the SIRIUS team, O. Cavani and B. Boizot, for operating electron irradiation SIRIUS facility at {\'E}cole Polytechnique [supported by the EMIR (R\'eseau national d’acc\'el\'erateurs pour les \'Etudes des Mat\'eriaux sous Irradiation) network, proposal 15-5788]. We thank Bing Li, W. R. Meier, and D. S. Robinson for experimental  support of the x-ray diffraction study.  The x-ray study used resources of the Advanced Photon Source, a U.S. Department of Energy (DOE) Office of Science User Facility operated for the DOE Office of Science
by Argonne National Laboratory under Contract No. DE-AC02-06CH11357.




\section{Supplementary information}

This article is published in Nature Communications \textbf{9}, 2786 (2018).
https://doi.org/10.1038/s41467-018-05153-0.

\begin{figure*}[htb]
\includegraphics[width=10cm]{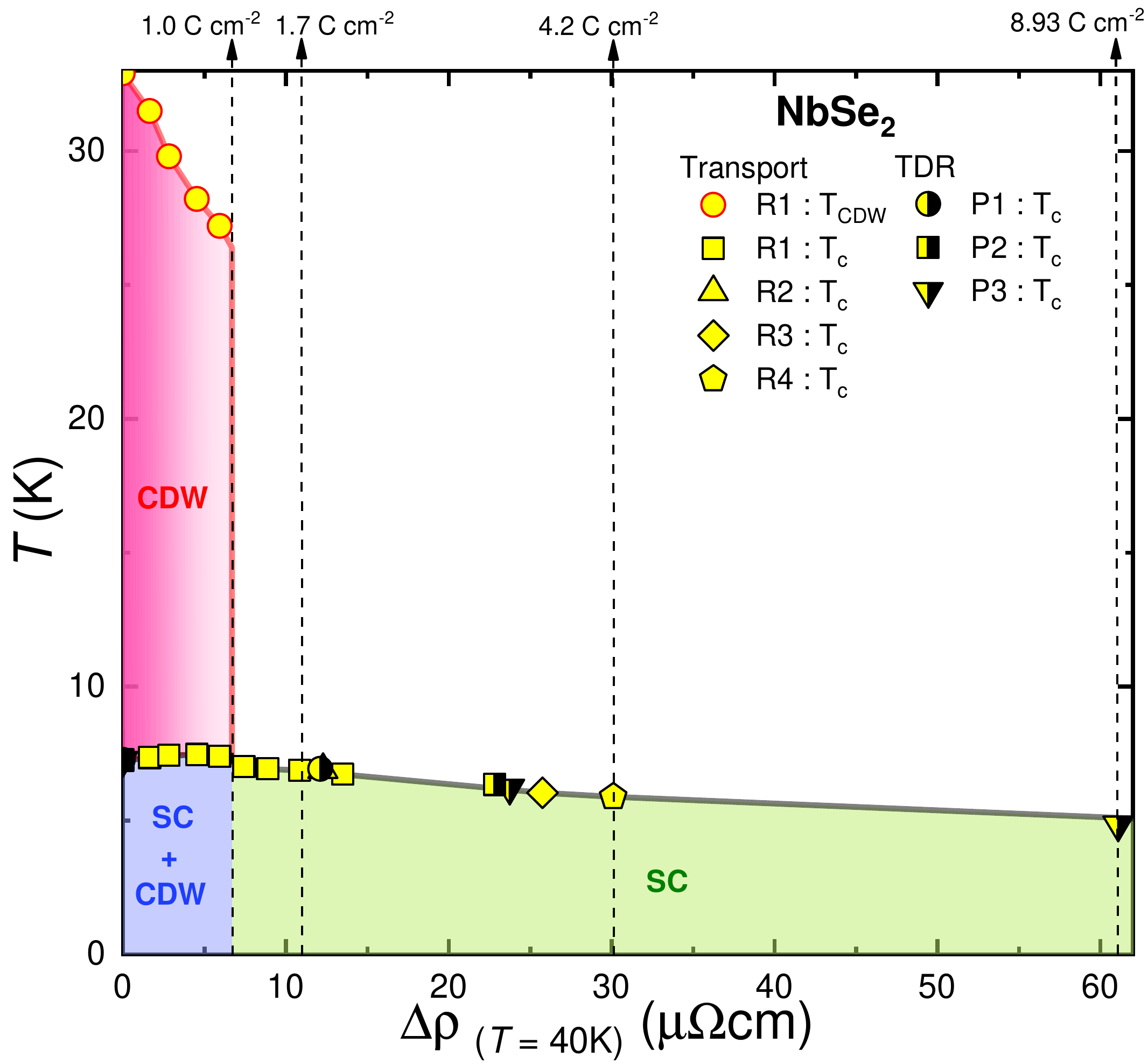}
\caption{\textbf{Full - scale version of the phase diagram shown in Fig.~4}. The highest dose is 8.93 C cm$^{-2}$ ($\Delta \rho_{(T = 40 K)}$ = 61 $\mu \Omega$cm).}
\label{fig_s1}
\end{figure*}

\begin{figure*}[htb]
\includegraphics[width=10cm]{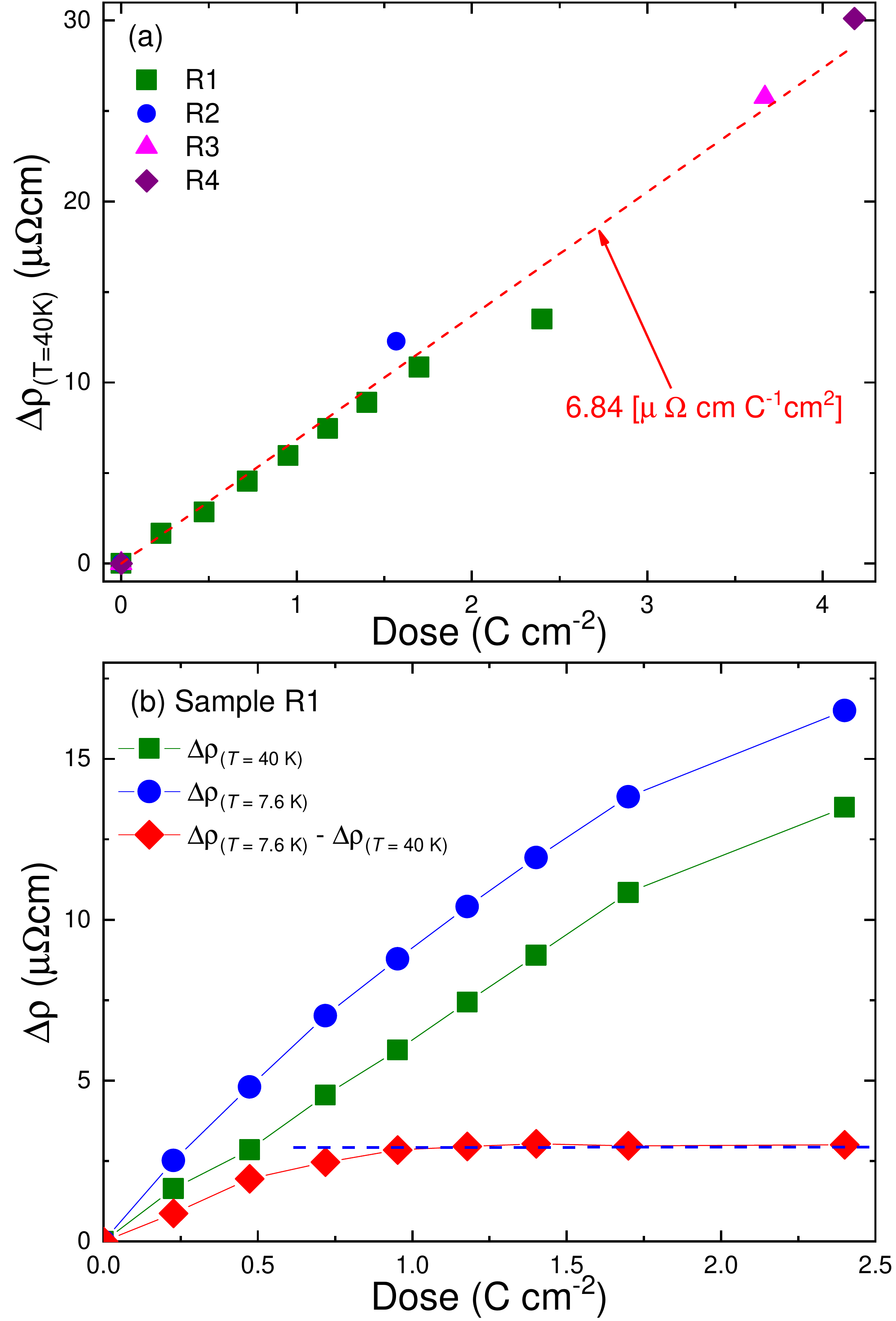}
\caption{\textbf{Relation between dose (C cm$^{-2}$) and $\Delta \rho_{(T = 40 K)}$ for sample R1.} (a) Dose (C cm$^{-2}$) versus $\Delta \rho_{(T = 40 K)}$. (b) Comparison between $\Delta \rho_{(T = 40 K)}$ and $\Delta \rho_{(T = 7.6 K)}$ that shows the suppression of the CDW above 1.0 C cm$^{-2}$. Refer to Fig.~2 (c).}
\label{fig_s2}
\end{figure*}

\end{document}